\documentclass{article} 
\usepackage[english]{babel} 
\usepackage{times}
\usepackage[iso]{umlaute} 
\usepackage{epsfig} 
\usepackage{doublespace}
\usepackage{amsmath}
\usepackage{fancyheadings}
\begin{document} 
\title{A study of transverse ply cracking using a discrete element method} 

\author{Falk K. Wittel$^1$ \thanks{Corresponding author, email adress: wittel@isd.uni-stuttgart.de}, Ferenc Kun$^2$, Bernd-H. Kröplin$^1$, Hans J. Herrmann$^3$ } 

\maketitle
\thispagestyle{myheadings} \markright{Submitted to the {\em Journal of Materials Science}} \headrulewidth 0pt
\begin{center}
$^1$ Institute for statics and dynamics of aerospace structures, University of Stuttgart,\\ Pfaffenwaldring 27, 70569 Stuttgart, Germany\\
$^2$ Department of Theoretical Physics, University of Debrecen, P.O.Box: 5,\\ H-4010 Debrecen, Hungary\\
$^3$ Institute for Computer Physics (ICA 1), University of Stuttgart, Pfaffenwaldring 27,\\ D-70569 Stuttgart, Germany 
\end{center}

\newpage

\begin{abstract}
We study the transverse cracking of the $90^\circ$ ply in $[0/90]_S$ cross-ply laminates by means of a discrete element method. To model the $90^\circ$ ply a two-dimensional triangular lattice of springs is constructed where nodes of the lattice model fibers, and springs with random breaking thresholds represent the disordered matrix material in between. The spring-lattice is coupled by interface springs to two rigid bars which represent the two $0^\circ$ plies in the model, which could be sublaminate as well. Molecular dynamics simulation is used to follow the time evolution of the model system. It was found that under gradual loading of the specimen, after some distributed cracking, segmentation cracks occur in the $90^\circ$ ply which then develop into a saturated state where the ply cannot support additional load. The stress distribution between two neighboring segmentation cracks was determined, furthermore, the dependence of the microstructure of damage on the ply's thickness was also studied. To give a quantitative characterization of stiffness degradation, the Young modulus of the system is monitored as a function of the density of segmentation cracks. The results of the simulations are in satisfactory agreement with experimental findings and with the results of analytic calculations.
\end{abstract}

\newpage
\section{Introduction} 
Transverse cracks of $90^\circ$ plies in $[0/90]_S$ cross-ply laminates \cite{tsai-hahn-80} appear early under tensile loading and continue forming until reaching a saturation state. The understanding of multiple transverse cracking is not only a phenomenon of scientific interest but also of practical importance. Transverse cracks can induce other damage mechanisms such as micro-delaminations followed by micro-buckling \cite{sjoegren-berglund-2000}. Failure generally occurs by the growth of nucleated or intrinsic defects on length scales much smaller than the sample size or representative volume element size used for the description of transverse cracking. One goal in materials design is the production of materials with flaw-insensitive or damage tolerant behavior, in a way that small flaws propagate in a stable manner up to global material failure \cite{krajcinovic-2000}. This necessitates the activation of interacting stress releasing mechanisms like crack deflections out of the preferred crack growth plane, micro-delaminations, induced microcracks or crack bridging. In any case the crack no longer propagates inside a homogeneous continuum, and inhomogeneity or disorder on certain length scales becomes a necessary element for the failure prediction of the composite. Models for the fracture simulation should allow the study of the growth of small cracks up to macroscopic size. 

For the last 20 years several analytical solutions were proposed for the calculation of the density of micro cracks under mechanical and thermal loading, using a variety of different approaches for stress analysis and different failure models \cite{nairn-hu-bark-93,nairn-hu-94,nairn-2000}. These models were combined with interacting damage mechanisms within a micro-mechanics of damage analysis to take micro-delaminations into account \cite{nairn-hu-92}. For woven fabric simple models were derived from cross-ply laminates, showing the same types of damage also found in cross-ply laminates. Even though analytic models are in good agreement with experimental findings, the description of the evolution of transverse cracks from the early nucleation of micro-cracks over coalescence and crack growth lies beyond the scope of an analytic approach.

Numerical simulations for the failure of the transverse ply have been performed at different length scales ranging from the microscopic scale, modeling single fibers \cite{weihe-etal-94,weihe-kroeplin-93/2} and models for small fiber clusters \cite{weihe-kroeplin-93,zhu-achenbach-91,asp-berglund-talreja-96} up to the meso scale studying homogenized plies \cite{lebon-baxevanakis-etall-98} with a variety of model approaches. Either symmetric fiber arrays built of unit cells consisting of a part of a fiber and matrix material are calculated using finite element methods \cite{zhu-achenbach-91,asp-berglund-talreja-96} or small clusters of fibers with microstructural disorder, embedded in matrix material are modeled directly, employing fictitious or discrete crack models \cite{weihe-kroeplin-93}. Unfortunately, these simulations are very time consuming and the results have limited meaning for the simulation of multiple cracking in thicker transverse plies. In addition, they do not allow predictions about size scaling or the possibility to model the dynamic fracture process itself. 

In the present paper we study the transverse cracking of the $90^\circ$ ply in a $[0/90]_S$ cross-ply laminate by means of a discrete element method. In order to model the $90^\circ$ ply, a two-dimensional triangular lattice of springs is constructed. The nodes of the lattice model fibers perpendicular to the plane of the lattice, and springs represent the matrix material in between. The spring-lattice is coupled by interface springs to two rigid bars which represent the two $0^\circ$ plies in the model, but could be adjacent sublaminates as well. Disorder is introduced by assigning randomly distributed breaking thresholds to the springs, {\it i.e.} a spring breaks if the load on it exceeds its breaking threshold. To simulate transverse cracking under uniaxial loading of the composite, external load is imposed on the triangular lattice by horizontally displacing the two edges and the upper and lower interface elements of the lattice. This corresponds to a load imposed by coupled flexural rigid bars along the load direction. The time evolution of the system is followed by solving the equation of motion of the nodes (molecular dynamics). With the discrete element method used in this study, relatively large system sizes can be handled, enabling simulation of the fracture processes with crack-crack interactions of thousands of cracks. Multiple cracking has been studied before, using random spring networks with springs of random distribution of strain failure \cite{curtin-scher-90,herrmann-roux-90,schlangen-garboczi-96,schlangen-garboczi-97}, but rarely in the field of fiber composites research. This approach is particularly suited for studies on dynamic instability in crack propagation, the collective behavior of many interacting cracks and size effects of multiple transverse cracking. Therefore, the discrete element method is practical for studying general features of the statics and dynamics of fracturing like the crack morphology, global fracture patterns due to the collective behavior of many interacting cracks as well as the dynamic instability in the propagation of cracks. Our investigation is focused on the process of damaging, stiffness degradation, micostructure of damage, and furthermore, on the relative importance of damage mechanisms like segmentation and delamination in the degradation process.

The paper is organized as follows: Section 2 gives a detailed description of the model construction. Tests of the model and the numerical results on transverse cracking obtained by simulations are presented in Section 3. An analytic approach worked out in the literature to the transverse cracking problem is summarized in Section 4. The simulational results are confronted with the analytic calculations and experimental results at the presentation of the analytic methods and also in the discussion part.

\section{The discrete element model} 
In order to study the formation of segmentation cracks, a two dimensional triangular spring network model of composites containing undeformable cells, is worked out. Molecular Dynamics (MD) simulation is used to follow the motion of each cell by solving Newton's equations of motion. In the present study we use a $4^{th}$ order Gear Predictor Corrector scheme \cite{allen-tildesley-87}. A general overview of MD simulations applied to composite materials can be found in \cite{herrmann-roux-90}.
 
With this method already small cracks are sharply defined, with the possibility to simulate simultaneously a conglomerate of cracks within rather large lattice sizes. The fundamental advantage of the lattice model used in this investigation is due to its simplicity, giving direct access and possible physical interpretation to each step of the algorithm. Consequently, one can modify the rules for features of interest, like the characteristic properties of size, strength or force in a rather straightforward and transparent way.
 
Despite of the advantages of lattice models, one has to be aware of the limitations of discrete network models for the simulation of failure. One type of questions deals with the abstraction of the underlying material regions with discrete elements concerning the attributes of the elements and their organization in a network as well as the boundary conditions and numerical approaches utilized. But these questions are independent from limitations imposed by the nature of fracture simulations with discrete elements. Discreteness implies, that the stresses in an element correspond to an average of the local stresses over the volume represented by this element. It is, therefore, impossible to model cases with stress transfer from failed material regions to the neighboring ones on a length scale smaller than the lattice spacing.
 
The model is composed in three major steps, namely, {\em a)} the implementation of the microscopic structure of the solid, {\em b)} the determination of the constitutive behavior, and finally {\em c)} the breaking of the solid.  
 
\paragraph{\em (a) Microstructure} 

The model is composed of circular cells of identical radius $r_f$ with their centers located on the nodes of a regular triangular lattice built out of equilateral triangles with $s$ being the characteristic lattice spacing as illustrated in Fig. 1. The cells represent the cross sections of fibers. The fiber radius $r_f$ depends on the fiber volume fraction $v_f$ in the transverse ply, and is adjusted following the expression 
\begin{alignat}{2}\label{Kantenlange} 
r_f=\frac{s\sqrt{v_f\sqrt{3}}}{\sqrt{2\pi}}, \qquad  0\leq v_f\leq  v_{f_{max}}=0.906.  
\end{alignat} 

The cell centers are connected by linear elastic springs with stiffness $E_s$, representing the material region with fiber, fiber matrix interface and embedding matrix material. For simplification purposes, the springs behave in a time independent way, but time dependent spring properties are easy to introduce. In our simulation, the cells are the smallest particles interacting elastically with each other. We use a two-dimensional simulation with cells of unit thickness, allowing only motion in the observation plane, with the two degrees of freedom being the two coordinates of the position of the cell center of mass. The utilization of a regular lattice is a clear neglect of the topological disorder, symptomatic for realistic fiber composite systems. 
Since we have no particular basis for a topological disorder, we make a computational simplification by assuming the disorder in spring properties to adequately account for all the relevant disorder present in the material. Disorder is given to the model by a Weibull distribution of breaking thresholds $F_d$ at the beginning of the simulation in the form
\begin{alignat}{2}\label{Weibull} 
P(F_d)=1-\exp\left[-\left(\frac{F_d}{F_0}\right)^m\right] 
\end{alignat} 
and is kept constant during the fracture process (quenched disorder). The Weibull modulus $m$ controls the degree of disorder in the distribution and it is usually chosen in the range $1.5\leq m \leq 20$, experimentally found to describe a variety of materials. $F_0$ represents the characteristic strength of the material. The top and bottom row of cells, as well as the left and right rows are connected to flexural rigid beams via interface springs, representing the $0^\circ-90^\circ$ interface of the composite. In the following $nx$ denotes the number of cells along the loading direction, and $ny$ along the perpendicular one.

\paragraph{\em  (b) Constitutive behavior} 
 
The fibers are assumed to be rigid undeformable cylindrical bodies with the possibility to overlap when they have contact. This overlap represents to some extent the local deformation of the fibers (soft particle approximation \cite{herrmann-hovi-luding-98}). A contact law is used to estimate the contact force $\vec{F}_{ij}^c$ between two circular cells $i$ and $j$ as a function of the overlap area $A$ (see Fig. 1$b$). The overlap area $A$ can be expressed in terms of the distance $r_{ij}$ of the centers as
\begin{equation} \label{overlaparea} 
A=\frac{4}{3}(2r_f-r_{ij})\sqrt{r_f^2-\frac{r_{ij}^2}{4}},
\end{equation} 
and the repulsive contact force $\vec{F}_{ij}^c$ between two cells is given by
\begin{equation} \label{contactforce} 
\overrightarrow{F}_{ij}^c=-\frac{E_{ft}A}{L_c}\overrightarrow{n_{ij}},
\end{equation} 
where $L_c$ is a characteristic length chosen to be equal to the lattice spacing $s$, $E_{ft}$ denotes the elastic modulus of a fiber perpendicular to the fiber direction, and $\vec{n}_{ij}$ is the unit vector pointing from particle $i$ to $j$. The contact law inside the system is only applied between two circular particles after the breaking of the connecting spring, preventing penetration of broken parts into each other. It is a drawback of this method that the overlap force $(\ref{contactforce})$ does not have a Hamiltonian formulation which implies that no energy can be associated to the deformation modeled by the overlap. The force $\vec{F}_{i}$ on the cell $i$ is the residual force of all attractive and repulsive forces on the cell calculated from their interaction with neighboring cells either by contact or by the springs. In principle, stress can be applied to the system either by changing the elastic modulus, the initial length of the springs or by applying a force on the interface elements. In the present paper for the purpose of the study of transverse ply cracking, mechanical load is introduced on the transverse ply by straining the rigid upper and lower beams in $x$-direction and moving the left and right beam, respectively (see also Fig. 1$a$).

\paragraph{\em  (c) Breaking of the solid} 
 
Stresses inside the system can be released either by the formation of surfaces (cracks) inside the transverse ply or by fracture of the interface elements on the boundary. In the framework of Discrete Element Methods, the complicated crack-crack interaction is naturally taken into account. If the total force on a spring $F$ exceeds its breaking threshold $F_d$ (damage threshold), its stiffness is abruptly reduced to zero, resulting in a load redistribution to neighboring springs in the next iteration steps. If after some iterations the neighboring springs exceed their threshold value due to the additional load, they fail too, giving rise to crack growth. Like in the bulk material, the interface springs break once their maximum load is exceeded. Their failure is interpreted as micro delamination. The breaking characteristics of the interface elements are in principle independent of those of the bulk springs. In the model this is captured by assigning damage thresholds to the interface springs using a Weibull distribution but with parameters different from the ones used for the bulk.

The time evolution of the particle assembly is followed by solving the equation of motion of the nodes ({\it i.e.} transverse fibers) 
\begin{eqnarray} 
  \label{eq:eom} 
  m\ddot{\vec{r}_i} = \vec{F}_i, \qquad i=1, \ldots , N 
\end{eqnarray} 
where $N$ denotes the number of nodes (fibers of the transverse ply), and $\vec{F}_i$ is the total force acting on node $i$. A $4^{th}$ order Predictor Corrector algorithm is used in the simulations to solve numerically the second order differential equation system Eq.\ (\ref{eq:eom}). After each integration step the breaking condition is evaluated for all the intact springs. Those springs which fulfill this condition are removed from the further calculations (spring breaking).

\section{Computer simulations} 
Before applying the model to study transverse cracking in fiber composites a variety of simulations have been performed in order to test the behavior of the model and to make parameter identification with respect to experiments. Therefore, reasonable characteristic values for the lattice spacing $L_c$, spring stiffness $E_s$ and strength $\sigma_c$, contact modulus $E_{ft}$ and time have to be chosen. The Young modulus and Poisson ratio of the model system was measured numerically by simulating uniaxial loading of a rectangular sample when springs were not allowed to break. To avoid the disturbing effect of elastic waves that can be induced by the loading in our dynamical model, the numerical experiment is performed in the way developed in \cite{kun-herrmann-96}: The two opposite boundaries of the solid are moved with zero initial velocity but non-zero acceleration. When a certain velocity is reached the acceleration is set to zero and the velocity of the boundaries is kept fixed, which ensures a constant strain rate loading. With the help of this slow loading the vibrations of the solid can be reduced  drastically compared to the case when the boundaries start to move with non-zero initial velocity. An additional way to suppress artificial vibrations in dynamical models, used in this study, is to introduce a small dissipation (friction or damping) for the springs. This dissipation has to be small enough not to affect the quasi-static results. In order to measure the macroscopic elastic behavior of the model solid, the total force needed to maintain the external constraint is monitored from which the stress $\sigma$ can be determined as a function of strain $\varepsilon$. A representative example of the macroscopic elastic behavior is shown in Fig. 2$a$, where a good agreement of the numerical and analytical behavior can be observed.

As a further test of the model, at a certain load level two parallel cracks were introduced in the lattice perpendicular to the loading direction spanning through the whole cross section of the transverse ply. Keeping fixed the external load, the distribution of the stress components along the interface between the transverse and the longitudinal plies was calculated. The distribution of $\sigma_{xx}$ is presented in Fig. 2$b$.

\subsection{Crack formation under gradual loading} 
The discrete element method gives the possibility to monitor the development of the microscopic damage in the specimen. Simulations showed that in accordance with experiments the damage process of the transverse ply is composed of two distinct parts. At the beginning of loading the weakest  springs break in an uncorrelated fashion when they get over-stressed, generating microcracks in the specimen. This primary uncorrelated microcrack nucleation is dominated by the disorder of the system introduced for the damage thresholds of the springs. The microcracks substantially change the local stress distribution in the ply leading to stress concentrations around failed regions, which gives rise to correlated growth of cracks under further loading, see Fig. 3$a$. Crack growth occurs perpendicular to the loading direction. Growing cracks or cracks formed by the coalescence of growing cracks can span the entire thickness of the transverse ply, resulting in segmentation, {\it i.e.} break-up of the ply into pieces, as is illustrated in Fig. 3$b$. Reaching the ply interface the crack stops without the possibility of penetration into the longitudinal ply in the model, as was observed in most of the experiments, but with the possibility of deflection at the ply interface. The simulated fracture process is similar to observations described in the literature \cite{sjoegren-berglund-2000}.

Further segmentation cracks mainly form between existing cracks until the crack density is saturated due to occurring micro delaminations. Micro delaminations usually start to occur when the crack density has already reached high values, depending on the thickness of the transverse ply. Fig. 3$d$,$e$ show that as a result of the formation of cracks spanning the entire thickness, the ply gets segmented into pieces, with delamination zones extending along the interfaces.

In order to obtain a quantitative characterization of the microstructure of damage, we analysed the spatial distribution of microcracks, {\it i.e.} the distribution of broken springs. A crack is identified as a connected set of broken springs in the triangular lattice taking into account solely nearest neighbor connections. A crack is considered to be a segmentation crack if it spans from one side of the transverse ply to the other one. Since cracks forming along the interface of the plies complicate the identification of segmentation cracks in the framework of our algorithm, we use now the inner part of the specimen for the cluster identification. The method is illustrated in Fig. 4 for two different snapshots of the loading process. Fig. 4$a$ and Fig. 4$c$ contain all the broken springs which are present at the given time, while in Fig. 4$b$ and in Fig. 4$d$ only the segmentation cracks are presented which are identified by our analysis. It can be observed that the number of microcracks and also the number of segmentation cracks increases with time and the segmentation cracks have a more or less quasi-periodic spacing. The formation of the segmentation cracks is very fast and often shows stable crack growth only during the crack initiation. After the formation of the first segmentation crack the specimen is strained further, keeping the velocity of the boundary cells constant, till enough energy is built up to activate the formation of the next segmentation cracks approximately midway between existing segmentation cracks.
 
The total number of microcracks divided by the total number of bonds in the system $N_{cracks}/N_{bonds}$ and the number of segmentation cracks $N_{seg}$ are presented in Fig. 5 as a function of strain $E_s\varepsilon/\sigma_c$ for several different values of the Weibull-modulus $m$.  

It can be observed that cracking initiates at a finite strain value $\varepsilon_{in}$, called damage initiation strain, and starting from this point $N_{cracks}$ monotonically increases during the entire loading process. Segmentation first occurs at a strain larger than $\varepsilon_{in}$ after the number of microcracks reached a certain value. The number of segmentation cracks $N_{seg}$  also increases, however, it gets saturated for larger strains in accordance with experimental observations.
 
Fig. 5 provides also an insight into the role of disorder in the damage process. Increasing the value of $m$, {\it i.e.} making the Weibull distribution narrower, which implies that the system gets more and more brittle, the damage initiation strain $\varepsilon_{in}$ and the corresponding strain value at which the segmentation starts increase, furthermore, the saturation value of the segmentation cracks also increases for more brittle systems. The slight decrease of $N_{seg}$ for wider breaking threshold distributions is an artifact of the cluster analysis that identifies segmentation cracks connected via large delamination zones as one cluster and therefore as one segmentation.

\subsection{Transverse ply stiffness degradation} 
All the observed forms of damage lead to a reduction of the total load portion, the transverse ply shares with the longitudinal plies. The ply stiffness is a macroscopic property, but the numerical model is of micro structural nature. Therefore, the different effects of damage on the global stiffness have to be obtained by calculating an effective stiffness of the transverse ply. In this study we calculate the effective Young's modulus $E_{eff}$ by cumulating the energy density $e_{pot}$ stored in the spring elements as
\begin{equation} 
E_{eff}=\frac{2\cdot e_{pot}}{\varepsilon^2}.
\end{equation} 
This assumption is justified due to the way of loading, resulting in mainly elongated springs, and it is also ensured that only a small portion of the cell-cell contacts are closing in the late stage of the simulations. Fig. 6 shows stiffness degradations with varying Weibull modulus and with cracking artificially switched off after a certain number of iteration steps in the simulations. It can be observed that the more brittle the system is (i.e. the higher the Weibull modulus $m$ is), the longer is the horizontal plateau that precedes the decreasing part of the effective young modulus of ply group 1 $E^{(1)}_{eff}$, which starts at the crack initiation strain. For high strains all the curves with different $m$ values tend to a finite but very small stiffness which is reached when the cracking stops due to saturation.  When cracking is artificially switched off after a certain time, the effective stiffness remains constant due to the constant amount of damage as expected. This corresponds directly to a change in the secant stiffness, useful for continuum damage models.

\subsection{Micro-cracking in laminates with varying cross-ply thickness} 
The damage evolution in the transverse ply of $[0,90_n]_s$ laminates depends not only on the disorder, but also on the thickness of the transverse ply. It is observed experimentally that for thicker transverse plies, the saturation crack density is smaller than for thin plies, with delamination starting earlier during the loading. Both effects can be observed qualitatively in Fig. 7, two compare snapshots of the damage process with different thicknesses see also Fig. 3$a-e$.

\section{Analytical model} 
Popular analytical models for the composite damage analysis are continuum damage mechanical (CDM) or micro-mechanics of damage (MMD) models. In CDM, the mechanical properties are expressed as functions of damage, with the aim of finding the constitutive relations between different states of stress, strain and damage. The representation of all kinds of damage within one damage tensor, expressing the state of damage that may include a combination of matrix cracking, fiber breakage, micro delamination and ply delamination, shows the dependency and limitation of CDM from experimental and/or theoretical input. The main disadvantage of a CDM approach for this study is believed to be, that CDM does not use crack propagation theories based on fracture mechanics, which is necessary for a damage propagation description.  
  
MMD analysis is based on the individual prediction of the initiation and propagation of the various types of damage, including possible interactions between all identified relevant damage mechanisms. Such analysis consist of two steps, a stress analysis, and a damage analysis. The stress analysis has to be undertaken in the presence of observed damage. The stress analysis is completely independent from the postulation of a failure criterion for the initiation and evolution of damage, but of course not from the damage itself. An overview of MMD analysis for matrix micro cracking can be found in \cite{nairn-hu-94,nairn-2000}. Since the analytic solution is in good agreement with experimental results, we use it for comparison with results of this study \cite{nairn-hu-92,nairn-2000}.  
 
\paragraph{\em  Stress analysis}
The stress field in presence of micro-cracking damage is calculated for a unit cell with cracks located at each end (Fig. 8). The entire specimen is built by a sequence of unit cells. A two-dimensional analysis for the $x-z-$plane derived by Hashin \cite{hashin-85,hashin-86}, using a variational mechanics analysis is employed. This way, the characteristic violation of the boundary condition caused by a non-zero interfacial shear stress at the surface of the ply crack, which is symptomatic for an one-dimensional analysis can be avoided. Following \cite{hashin-85} we assume that the $x$-axis tensile stress in each ply group is independent of $z$, that ply cracks span the entire ply thickness and that delaminations are always symmetric \cite{nairn-hu-92}.

The x-axis tensile stress of a micro-cracked laminate can be derived with force balance 
\begin{alignat}{2}\label{sigmaxx} 
 \sigma^{(1)}_{xx}=E^{(1)}_x/E_c^0\sigma_0-\psi(x), \qquad
 \sigma^{(2)}_{xx}=E^{(2)}_x/E_c^0\sigma_0+\frac{\psi(x)}{\lambda},
\end{alignat} 
with the undetermined function $\psi(x)$ and $\lambda=t_2/t_1$ expressing stress perturbations caused by the segmentation cracks. $E^{(i)}_x$ is again the x-direction young's modulus of ply group $i$ and $0$ denotes global laminate properties. The integration of the stress equilibrium equations with the unit cell boundary conditions gives the shear and transverse stresses in the unit cell in terms of $\psi(x)$. The function $\psi(x)$ that minimizes the complementary energy gives the best approximation to the micro-cracked cross-ply laminate stress state. The solution of the Euler equation for finding $\psi(x)$ can be found in detail in \cite{hashin-85}. Fig.9 shows the stress distributions inside a unit cell for interfacial values of $\sigma_{xz}$ and $\sigma_{zz}$ as well as $\sigma_{xx}$ for longitudinal and transverse ply, with the comparison of $\sigma_{xx}^{(1)}$ to the stress calculation from our numeric simulation (compare Fig. 9$a$). A reasonable agreement of the analytic solution and of the corresponding simulation can be observed.
 
The composite properties for the laminate have been calculated by means of micro-mechanics from the bulk properties for the fiber and matrix constituents \cite{stellbrink-96,tsai-hahn-80}. For this work we renounce a detailed tabulation of bulk and composite material properties, since emphasis is given on a dimensionless presentation of results. Nevertheless, the characteristic length of the numerical model is scaled with the fiber volume fraction and an average fiber diameter of $7\mu m$, whenever it was necessary for comparative reasons.
 
\paragraph {\em  Failure model}

The failure criterion used in the analytical model, is an energy criterion \cite{nairn-89}, allowing a formation of a new crack, whenever the strain energy release rate (ERR) $G_m$ associated to the formation of a new micro crack exceeds some critical value $G_{mc}$, also called intralaminar fracture toughness or micro-cracking fracture toughness. The ERR is also calculated with Hashin's two-dimensional, variational mechanics analysis \cite{nairn-89}.
\begin{equation} \label{ERR} 
G_m=\sigma_{x0}^{(1)^2}t_1\sqrt{C_1(C_4-C_2+2\sqrt{C_1C_3}}. 
\end{equation} 
\begin{alignat}{2} 
C_1 &=\frac{1}{E_x^{(1)}}+\frac{1}{\lambda E_x^{(2)}}, &\qquad C_2 
&=\frac{\nu^{(1)}_{xz}}{E_x^{(1)}}\left(\lambda+\frac{2}{3}\right)-\frac{\lambda\nu_{xz}^{(2)}}{2E_x^{(2)}}, 
\\C_3&=\frac{1}{60E_z^{(1)}}(15\lambda^2+20\lambda+8)+\frac{\lambda^3}{20E_z^{(2)}}, 
&\qquad C_4 &=\frac{1}{3G_{xz}^{(1)}}+\frac{\lambda}{3G_{xz}^{(2)}}. 
\end{alignat} 

$E_{xx}^{(1)}$, etc, are the material properties of the plies in conventional notation. 

\paragraph{\em Delamination} 
 
Microcrack-induced delamination is a damage mechanism, that competes with the formation of new segmentation cracks in the transverse ply. Experimental and numerical studies show, that ply crack-induced delamination does not start before a critical crack density is reached.

The two-dimensional, variational mechanics analysis of \cite{nairn-hu-94} described above can be extended to account for delaminations emanating from ply crack tips. The assumption that $\sigma_{xx}^{(1)}$ is independent of $z$, indicates identical delamination lengths emanating from the top and bottom crack tips. This leads to a simple stress state for the delaminated regions:
\begin{alignat}{2}
   \sigma_{xx}^{(1)} &= 0, &\qquad \sigma_{xz}^{(1)} &=\sigma_{zz}^{(1)}=0,\\ 
   \sigma_{xx}^{(2)} &= 
   \sigma_{x0}^{(2)}+\frac{\sigma_{x0}^{(1)}}{\lambda}=\frac{1+\lambda}{\lambda}\sigma_0, &\qquad 
   \sigma_{xz}^{(2)} &=\sigma_{zz}^{(2)}=0. 
\end{alignat} 
In the region between the delaminations, the stress distribution is equal to that existing between two microcracks with the dimensionless half spacing $\rho-\delta$ with $\delta=(d1+d2)/2t_1$ and $\rho=a/t_1$. The ERR for the initiation of delaminations in the unit cell is given by:
\begin{equation} 
G_d=C_3t_1\left(\frac{E_x^{(1)}}{E_c}\sigma_0\right)\frac{2\chi'(0)-\chi'(\rho_k)}{2}, 
\end{equation} 
$\chi(\rho)$ is a function that corresponds to the excess strain energy in a unit cell of damage caused by the presence of ply cracks \cite{hashin-85,hashin-86}. Assuming that the critical energy release rate for delamination $G_{mc}=G_{dc}$, we compare the dimensionless ERR as a function of the crack density to find the crack density for the initiation of delaminations \cite{nairn-hu-92}. In the analytic framework, with the onset of delamination, no further segmentation takes place, so the crack density for delamination onset is equal to the crack density of the characteristic damage state (CDS). In Fig. 9$b$ the analytic values for the CDS as a function of transverse ply thickness are compared to our simulation results and to some limited experimental data. Since our numerical model is more realistic in the sense that it captures more details of the microscopic damage mechanisms, a satisfactory agreement can be stated between the simulation and the experimental findings, however, the analytic curve falls a bit further from them.

The reduction of the transverse ply stiffness is compared to simulation results in Fig. 10. It is obvious, that the analytic model without delamination has no segmentation crack saturation, leading to a higher crack density. Numerical results for two different transverse ply thicknesses show qualitative agreement, only with the stiffness reduction being zero at the saturation crack density.

\section{Discussion}

We introduced a disordered spring network model to study the transverse cracking of the $90^\circ$ ply in $[0/90_n]_S$ cross-ply laminates. The main advantage of our modeling is that it naturally accounts for the complicated local stress fields formed around failed regions in materials, furthermore, it captures the gradual activation of the relevant failure mechanisms and their interactions during the fracture process.

We have demonstrated that our discrete element model provides a deep insight into the damage process occurring under gradual loading of cross ply laminates. Quantitative results have been obtained on the microstructure of damage, on the local stress distribution between cracks and on the degradation of the ply's stiffness.

The results obtained by numerical simulations have been confronted with experimental findings \cite{kim-aoki-83} and also with the analytic approach of Nairn and Hu. Good agreement was found between the numerical, analytical and experimental results. However, the numerical simulations proved to be more realistic than the simple analytic approaches, due to the more detailed description of microscopic damage mechanisms as built in the discrete model.

Despite of the realistic model construction our study makes some technical simplifications which have to be overcome in the future. For instance, due to the difficulties in the identification of the fracture properties of interface bonds between plies, no simulations have been performed varying the relative breaking thresholds of interface and bulk springs. To clarify this problem a detailed parameter study will be carried out in a forthcoming publication.

The capabilities of our model are not limited to uniaxial loading, it can be applied to study damage processes of the transverse ply in cross-ply laminates under varying loading cases including also the thermal degradation process occurring during the production process of fiber composites. Studies in this direction are in progress.

\newpage
\section{Acknowledgment} 
The presented work is partly funded by the German Science Foundation (DFG) within the Collaborative Research Center SFB 381 'Characterization of Damage Development in Composite Materials using Non-Destructive Test Methods' which is gratefully acknowledged, as well as the support under NATO grant PST.CLG.977311. F.\ Kun is grateful for financial support of the Alexander von Humboldt Stiftung (Roman Herzog Fellowship), and also for the B\'olyai J\'anos fellowship of the Hungarian Academy of Sciences.
\newpage
\bibliographystyle{/scratch3/paper/latex/spiebib.bst} 
\bibliography{/home/wittel/Dir_tex/bibtex/wittel} 
\newpage
\section{Figure Legends}
\begin{itemize}
\item{Figure 1: Microstructure of the model $(a)$ and cells in contact $(b)$} 
\item{Figure 2: $(a)$ Macroscopic constitutive behavior of the uncracked specimen in a dimensionless form, obtained by uniaxial loading (switching off the spring breaking in the simulation). $(b)$ Stress distribution $\sigma_{xx}/\sigma_{x0}$ between two cracks.}
\item{Figure 3: Snapshots of the model system with the size $nx=800$, $ny=10$. $(a)$ nucleated microcracks, distributed over the  whole test cell, and the formation of the first segmentation crack. $(b)$ Quasi periodic segmentation pattern with corresponding stress distribution $(c)$. $(d)$ Segmentation cracks and delaminations at crack density saturation state. $(e)$ The saturated state for a system of $nx=800$, $ny=50$}
 \item{Figure 4: Identification of segmentation cracks at two different stages of the damage process. $(a)$ and $(c)$ show all the microcracks which are present at a given state, in $(b)$ and $(d)$ only the identified segmentation cracks are presented.}
 \item{Figure 5: $(a)$ The total number of microcracks $N_{cracks}$ divided by the total number of bonds $N_{bonds}$, and $(b)$ the number of segmentation cracks $N_{seg}$ as a function of strain $E_s\varepsilon/\sigma_c$. Ply thickness $ny$ was 10 cells. Smooth curves were obtained by averaging over six samples.}
\item{Figure 6: Stiffness reduction with varying Weibull moduli $(a)$ and for $m=4$ switching off the breaking after a certain number of iteration steps.}
\item{Figure 7: $(a)$ The total number of microcracks $N_{cracks}$ divided by the total number of bonds $N_{bonds}$, and $(b)$ the number of segmentation cracks $N_{seg}$ as a function of strain $E_s\varepsilon/\sigma_c$ for systems of changing thickness $ny$.}
\item{Figure 8: Side view of an unit cell representing a cross ply laminate with segmentation cracks and delaminations.}
\item{Figure 9: Stress distribution inside the unit cell for $a=2t_1$ and $t_1=t_2$ $(a)$, and characteristic crack density, analytic results for $G_{c_{del.}}=G_{c_{seg}}$, numerical and experimental \cite{kim-aoki-83} results $(b)$.}
\item{Figure 10: Comparison of analytic predictions for the loss of stiffness $E_x^1/E_{x0}^1$ with simulation results by our model.}
\end{itemize}
\newpage
\begin{figure}[H] 
\epsfig{bbllx=62,bblly=602,bburx=552,bbury=802, 
file=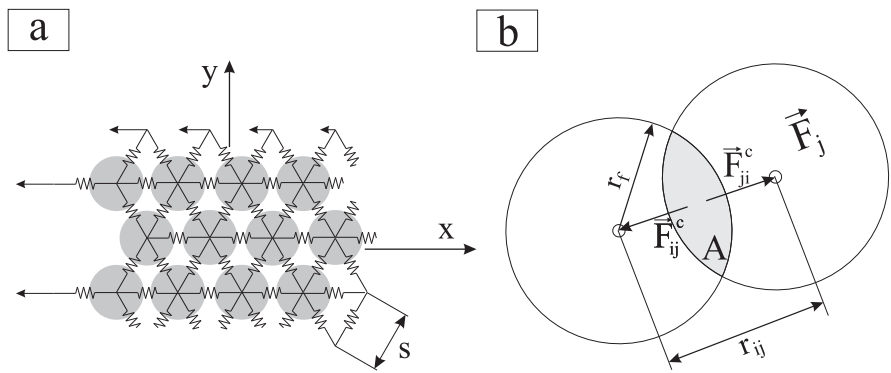, width=12cm} 
\label{microstructure}
\caption{Microstructure of the model $(a)$ and cells in contact $(b)$} 
\end{figure} 
\begin{figure} 
\begin{center} 
\begin{minipage}[t]{6.2cm} 
\epsfig{bbllx=11,bblly=20,bburx=400,bbury=400, 
file=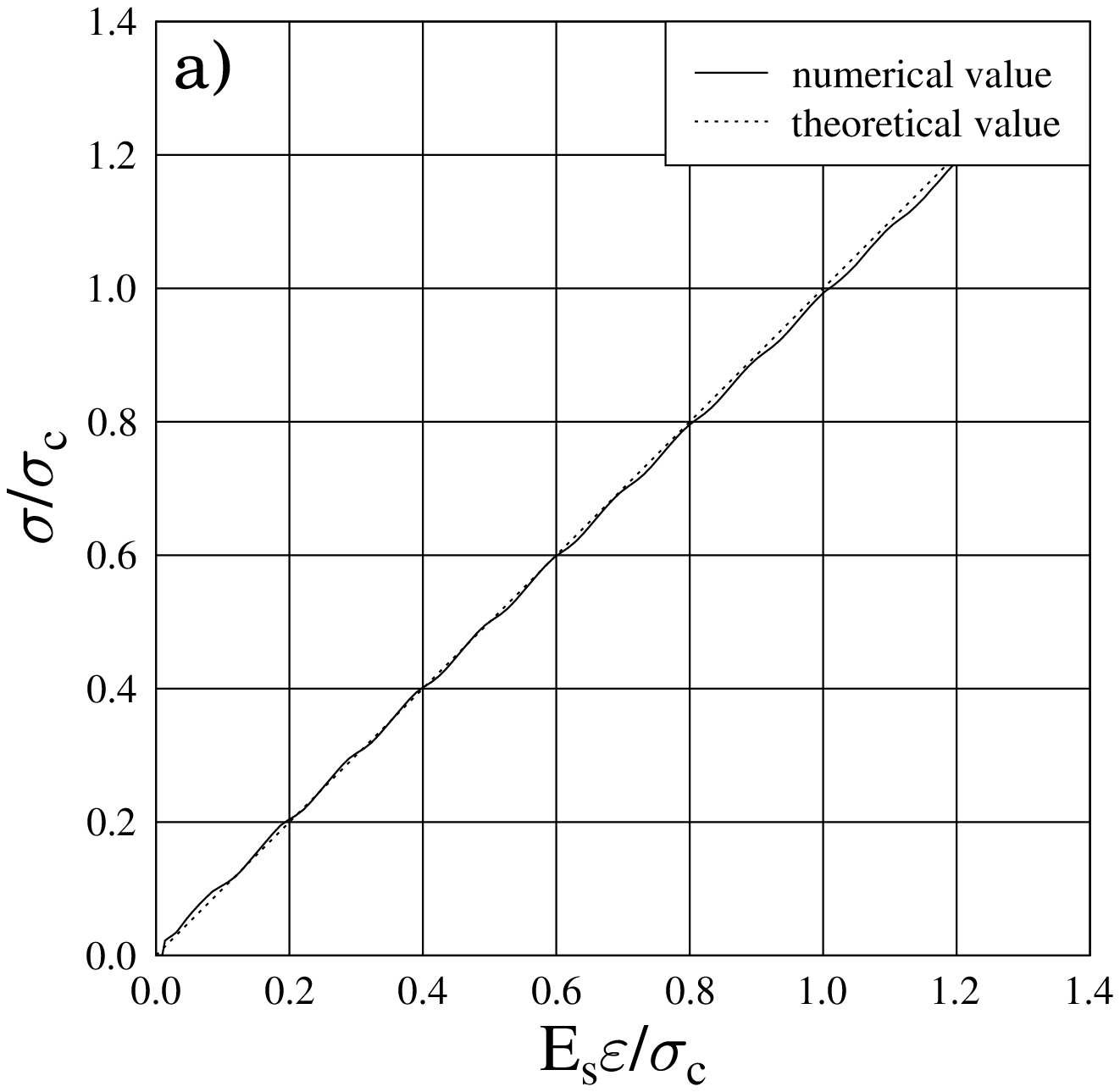, width=6.2cm} 
\end{minipage} 
\begin{minipage}[t]{6.2cm} 
\epsfig{bbllx=40,bblly=54,bburx=423,bbury=423, 
file=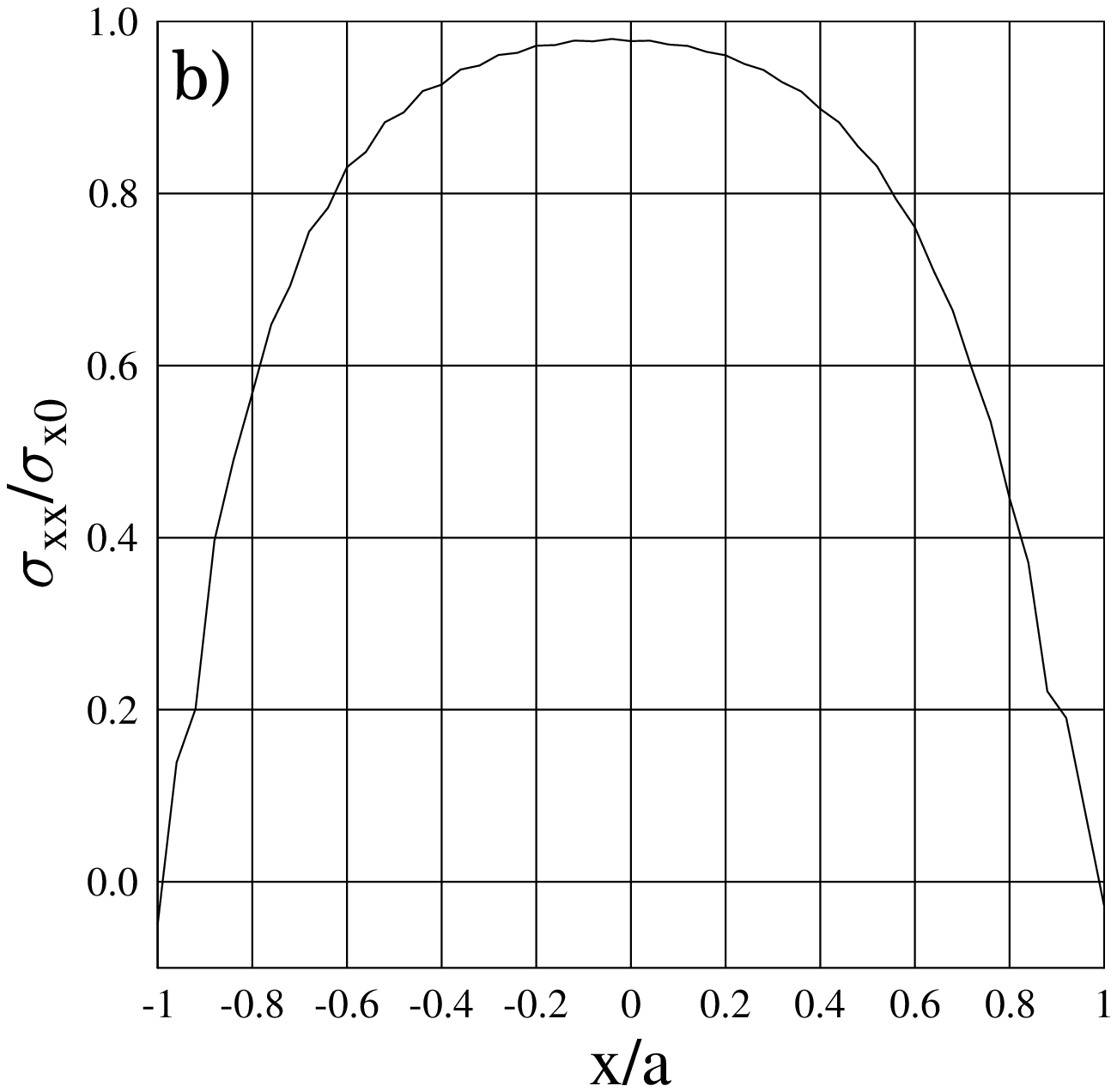, width=6.2cm} 
\end{minipage} 
\caption{$(a)$ Macroscopic constitutive behavior of the uncracked specimen in a dimensionless form, obtained by uniaxial loading (switching off the spring breaking in the simulation). $(b)$ Stress distribution $\sigma_{xx}/\sigma_{x0}$ between two cracks.}
\end{center} 
\label{fig:tests} 
\end{figure}   
\begin{figure}[H]
\begin{center} 
\epsfig{bbllx=-16,bblly=-280,bburx=1081,bbury=645, file=bilder/snapshots.eps, width=13cm} 
\end{center}
\caption{Snapshots of the model system with the size $nx=800$, $ny=10$. $(a)$ nucleated microcracks, distributed over the  whole test cell, and the formation of the first segmentation crack. $(b)$ Quasi periodic segmentation pattern with corresponding stress distribution $(c)$. $(d)$ Segmentation cracks and delaminations at crack density saturation state. $(e)$ The saturated state for a system of $nx=800$, $ny=50$}
\label{fig:snapshots}
\end{figure} 
\begin{figure}[H]
\begin{center}
\epsfig{bbllx=3,bblly=10,bburx=512,bbury=215, file=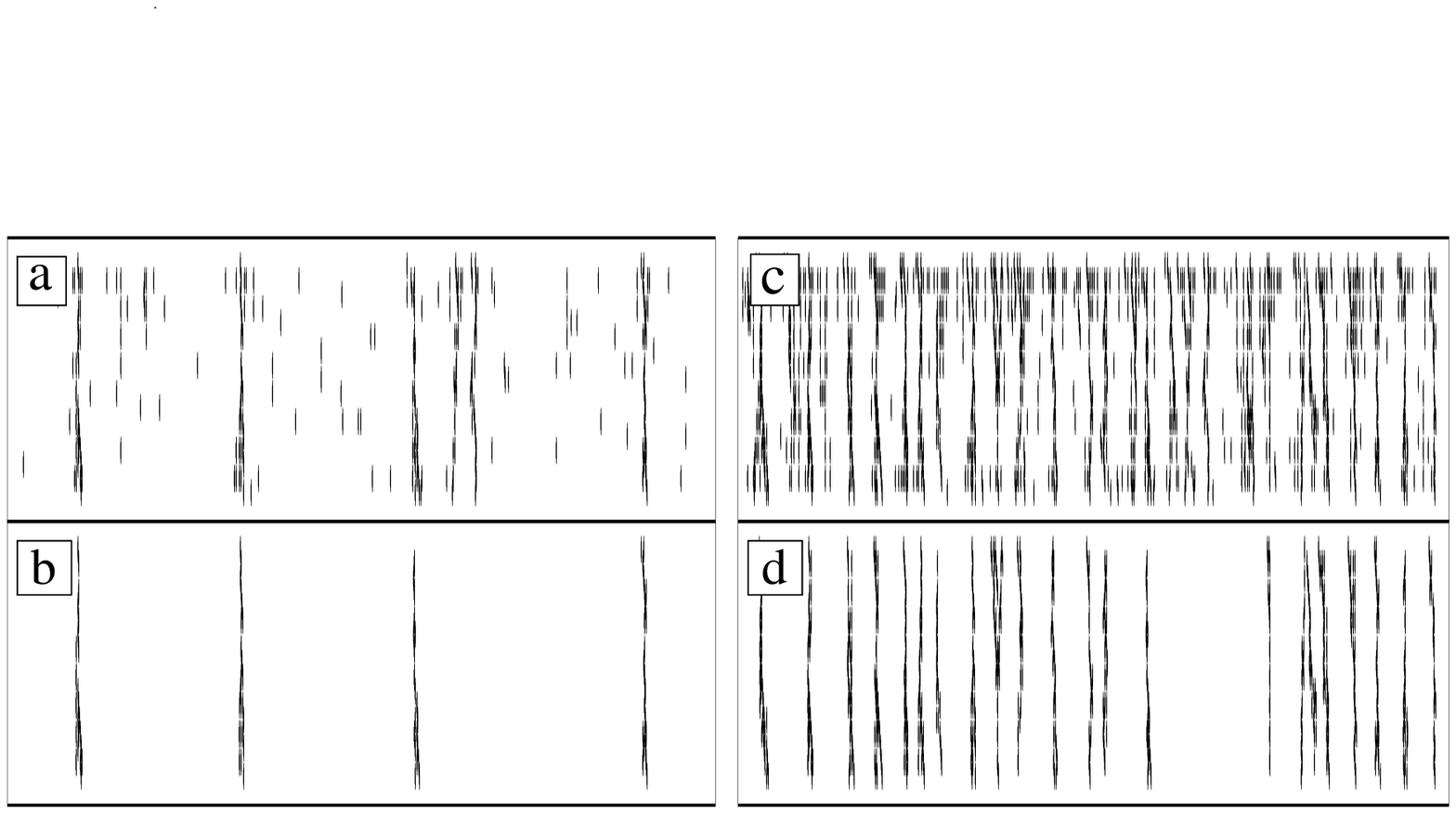, width=12cm}
 \caption{Identification of segmentation cracks at two different stages of the damage process. $(a)$ and $(c)$ show all the microcracks which are present at a given state, in $(b)$ and $(d)$ only the identified segmentation cracks are presented.}
\label{fig:cracks} 
\end{center} 
\end{figure}   
\begin{figure} [H]
\begin{center} 
\epsfig{bbllx=104,bblly=113,bburx=477,bbury=637, file=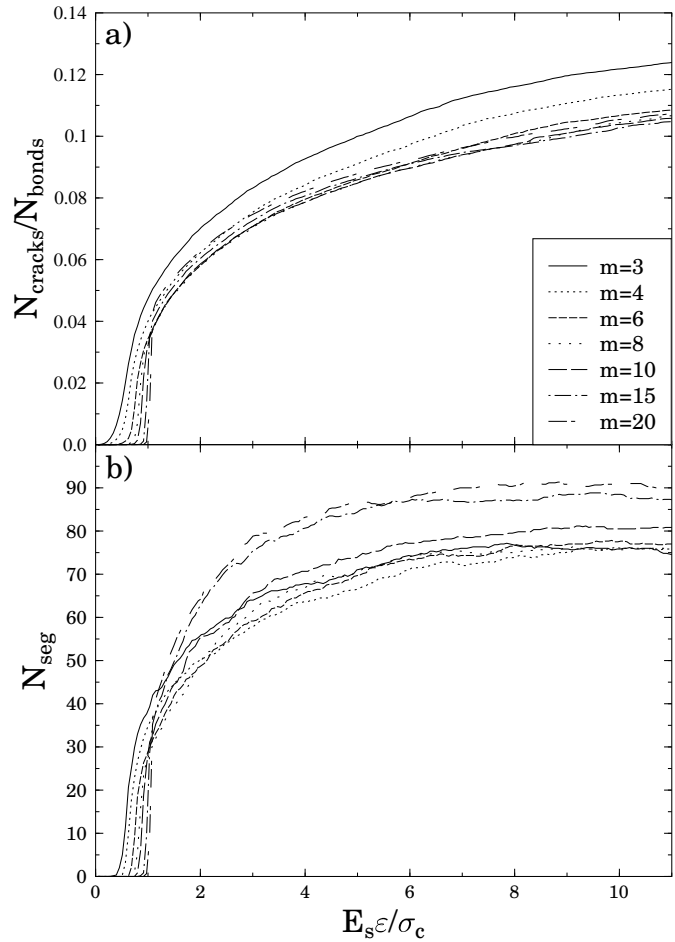, width=9cm} 
 \caption{$(a)$ The total number of microcracks $N_{cracks}$ divided by the total number of bonds $N_{bonds}$, and $(b)$ the number of segmentation cracks $N_{seg}$ as a function of strain $E_s\varepsilon/\sigma_c$. Ply thickness $ny$ was 10 cells. Smooth curves were obtained by averaging over six samples.}
\label{fig:crackdist} 
\end{center} 
\end{figure}   
\begin{figure}[H]
\begin{minipage}[t]{6.0cm} 
\epsfig{bbllx=30,bblly=47,bburx=420,bbury=420, file=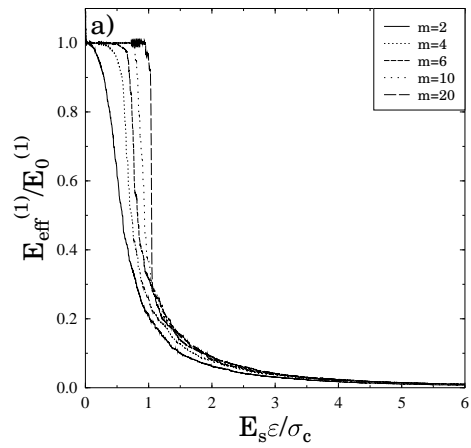, width=6.2cm} 
\end{minipage} 
\begin{minipage}[t]{6.0cm} 
\epsfig{bbllx=30,bblly=47,bburx=420,bbury=420, file=bilder/energy2.eps, width=6.2cm} 
\end{minipage} 
\label{fig:emodred} 
\caption{Stiffness reduction with varying Weibull moduli $(a)$ and for $m=4$ switching off the breaking after a certain number of iteration steps.}
\end{figure} 
\begin{figure} [H]
\begin{center} 
\epsfig{bbllx=100,bblly=113,bburx=477,bbury=637, file=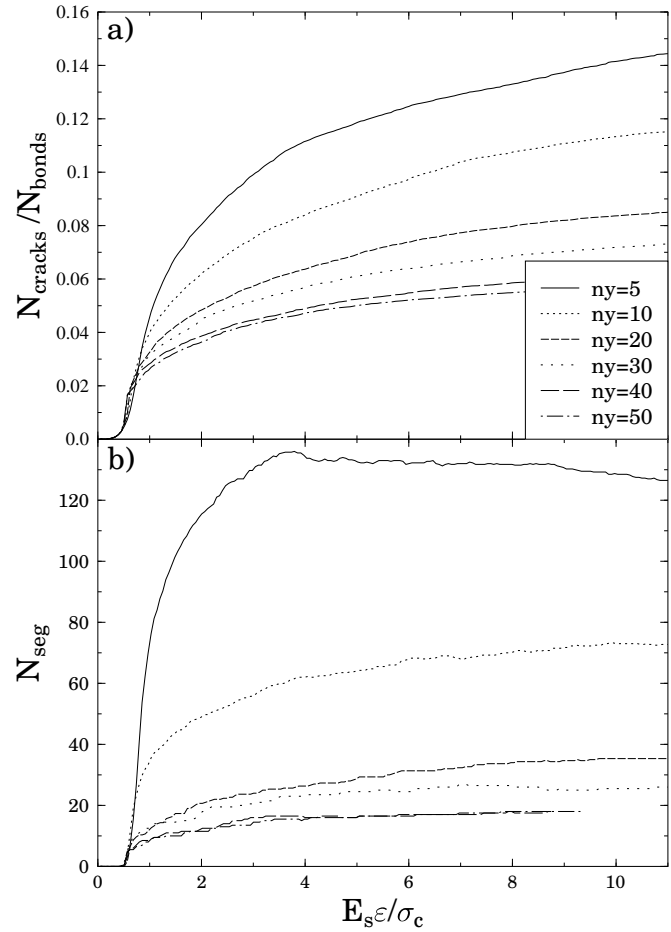, width=9cm} 
 \caption{$(a)$ The total number of microcracks $N_{cracks}$ divided by the total number of bonds $N_{bonds}$, and $(b)$ the number of segmentation cracks $N_{seg}$ as a function of strain $E_s\varepsilon/\sigma_c$ for systems of changing thickness $ny$.}
\label{fig:crackdist2} 
\end{center} 
\end{figure}    
\begin{figure}[H] \label{cdtransdef} 
\begin{center} 
\epsfig{bbllx=112,bblly=385,bburx=455,bbury=605, file=bilder/cdtransdef.eps, width=10cm} 
\end{center} 
\caption{Side view of an unit cell representing a cross ply laminate with segmentation cracks and delaminations.} 
\end{figure} 
\begin{figure}[H] 
\begin{minipage}[t]{6.2cm} 
\epsfig{bbllx=0,bblly=0,bburx=1471,bbury=2284, file=bilder/stress_nairn_paper.eps, width=6.2cm}
\end{minipage} 
\begin{minipage}[t]{6.2cm} 
\epsfig{bbllx=72,bblly=323,bburx=447,bbury=691, file=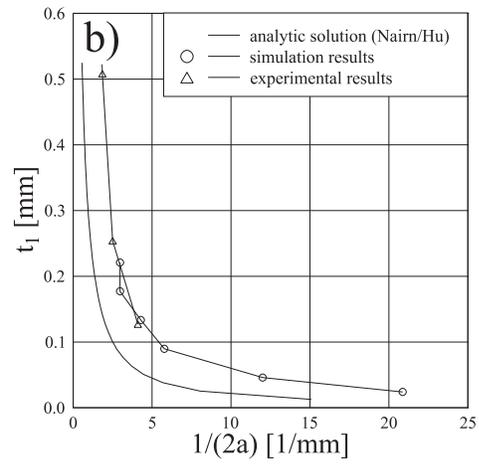, width=6.2cm} 
\end{minipage} 
\label{stressstate}
\caption{Stress distribution inside the unit cell for $a=2t_1$ and $t_1=t_2$ $(a)$, and characteristic crack density, analytic results for $G_{c_{del.}}=G_{c_{seg}}$, numerical and experimental \cite{kim-aoki-83} results $(b)$.}
\end{figure} 
\pagestyle{fancy} \markright{F. K. Wittel, F. Kun, B. H. Kröplin, H. J. Herrmann: {\em A 'Study of Transverse Ply Cracking Using a Discrete Element Method}} \headrulewidth 0pt
\begin{figure}[H] \label{stiffloss} 
\begin{center} 
\epsfig{bbllx=30,bblly=48,bburx=425,bbury=425, file=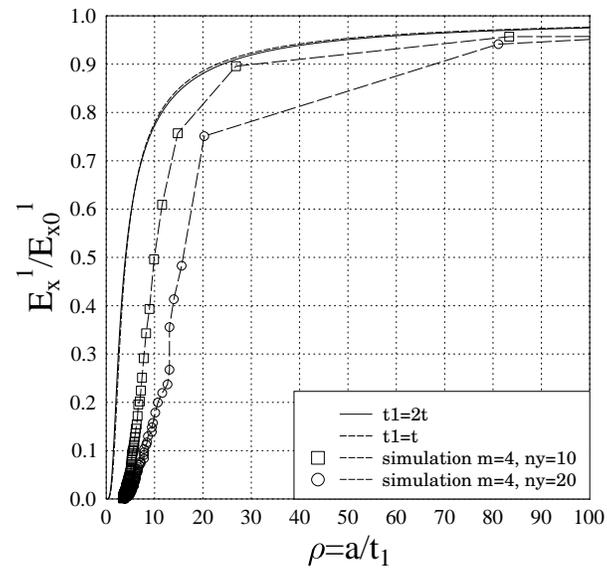, width=8cm} 
\end{center} 
\caption{Comparison of analytic predictions for the loss of stiffness $E_x^1/E_{x0}^1$ with simulation results by our model.} 
\end{figure} 

\end{document}